\documentclass[11pt]{article}
\usepackage[T1]{fontenc}
\usepackage{tgbonum}
\usepackage[utf8]{inputenc}
\usepackage{times}
\usepackage{geometry}
\usepackage[utf8]{inputenc}
\usepackage[T1]{fontenc}
\usepackage{enumitem,amssymb}
\usepackage{ragged2e}
\usepackage{dirtytalk}
\usepackage{amsmath}
\usepackage{graphicx}
\usepackage{comment}
\usepackage{eurosym}
\usepackage{multicol}
\usepackage[normalem]{ulem}
\usepackage[usenames]{xcolor} 
\usepackage[colorinlistoftodos]{todonotes}
\definecolor{xlinkcolor}{cmyk}{1,1,0,0}
\usepackage{url}
\usepackage[
 colorlinks=true,    
 linkcolor=xlinkcolor,     
 citecolor=xlinkcolor,     
 filecolor=xlinkcolor,  
 urlcolor=xlinkcolor,      
 final=true
]{hyperref}
\usepackage[super,sort&compress]{natbib}
\usepackage{enumitem}
\setenumerate{itemsep=0mm}
\usepackage{authblk}
\usepackage[belowskip=-15pt,aboveskip=-15pt]{caption}
\usepackage{subfiles}
\usepackage[nottoc]{tocbibind}

\usepackage{orcidlink}

\usepackage{footnote}
\usepackage{footmisc}

\newcounter{FPcomment}

\title{Snowmass Early Career\\The Key Initiatives Organization\\
\vspace{0.5cm}
\small{\scshape A Snowmass Contributed White Paper to the Community Engagement Frontier}
}

\author[1,2]{Joshua Barrow\thanks{jbarrow@fnal.gov}}
    \affil[1]{The Massachusetts Institute of Technology}
    \affil[2]{Tel Aviv University}
\author[3,4]{Kristi L. Engel\thanks{klengel@terpmail.umd.edu}}
    \affil[3]{University of Maryland, College Park}
    \affil[4]{Los Alamos National Laboratory}
\author[5]{Tiffany R. Lewis\orcidlink{0000-0002-9854-1432}\thanks{tiffanylewisphd@gmail.com}}
    \affil[5]{NASA Postdoctoral Fellow, Astroparticle Physics Lab, Goddard Space Flight Center}
\author[6]{Sara M. Simon\thanks{smsimon@fnal.gov}}
\author[7]{Jorge Torres}
    \affil[7]{Yale University}
    
\date{On the Behalf of Snowmass Early Career \\ \today}

\begin{document}

\maketitle
\vspace{-0.5cm}
\tableofcontents
\newpage
\section{The Participation of Early Career Members During Snowmass 2021}
In April 2020, the 2019 and 2020 American Physical Society's Division of Particles and Fields (APS DPF) Early Career Executive Committee (ECEC) members were tasked with ``organizing the formation of a representative body for High-Energy Physics (HEP) early career members for the Snowmass process" by the DPF Executive Committee. While during the previous Snowmass there was a group of early career members actively participating in the process, it was largely limited to those based at Fermilab and who were historically ``in the know"; thus, the DPF ECEC endeavored to gain broad community feedback at every step of initiating the Snowmass process. Here, we outline the structure we developed and the process we followed to help provide context and guidance for future early career Snowmass efforts. Hopefully some of our thoughts will help you to avoid the pitfalls we encountered. 

Throughout this document, please bear in mind that the 2021 Snowmass process took place almost completely during the Coronavirus-19 Pandemic, and while we recognize that this circumstance has had a huge effect on how this Snowmass has progressed, it may also be difficult for us to disentangle the two since none of us have experience with a Snowmass that is not during a pandemic, yet.  Culturally, this Snowmass was also organized in the wake of the Black Lives Matter movement, which thrust ideas about inclusion and equity into the national consciousness, especially in the context of amplifying marginalized voices. This informed the careful consideration of leadership structure and the focus on inclusively. As white papers were submitted and reports organized, Russia attacked Ukraine. This lead to last minute challenges in author and affiliation lists because there was a movement in the U.S. to protest the war by refusing to list Russian institutions on U.S. publications. To those who take up the next call, we wish you all the best, and look forward to working with you.

Throughout this document, the authors reflect on the triumphs and pitfalls of a program created from nothing over a very short period of time, by people with good intentions, who had no prior experience in building such an organization. Through this exercise of reflecting, we sometimes find that we would recommend a different path to our future selves. This should never be interpreted as a criticism of any of the individuals who gave their time and energy in earnest, to whom we are exceedingly grateful. Insomuch as there are things to find fault with, it is in the robustness of the systems we built and refined, and there is no subtext intended as we describe those systems. 

\section{Initial Formation and Plans}
First, we broadly advertised open nominations for early career leadership in the Snowmass process via email communiques to many major collaborations, the Snowmass email list, and placing an announcement in the APS DPF newsletter. The information was also shared with other APS Divisions as they are represented on the Snowmass Advisory Committee. Nominations began in April 2020 and were open until early June. Self-nominations were permitted. The time frame for nominations was initially set to one month (April), but was extended to a total of two months at the request of the DPF Executive Committee. Over 250 nominations were received, and it was ensured that there were nominees across all Frontiers. Town hall meetings open to all early career members were held in May and June to gather community input on the format and charge for Snowmass Early Career (SEC). 
Kick-off meetings of the general Snowmass Early Career (SEC) group and leadership hand-off to SEC initiative sub-groups started in late June. The primary points of feedback from these early meetings were:
\begin{itemize}
    \item A representative group for the early career HEP community is needed beyond the Snowmass process, and SEC is a good starting point to form a more permanent HEP early career organization
    \item An election process to down-select nominees was perceived as unfairly skewed to favor members of large experiments and large universities and could run contrary to the goal of a diverse and representative group
    \item Agreement that SEC representatives for each Snowmass Frontier would benefit the early career community and the Snowmass process
    \item Input on key initiatives that would form the core structure of SEC.
\end{itemize}

One important step was defining who fell under the umbrella of early career. We wanted to be inclusive of people on alternative career paths and those who had career gaps, so we agreed on the following definition of early career:

\vspace{0.2cm}
{\footnotesize\noindent\say{As a guideline, SEC roughly defines early career as up to $\sim10$ years post-PhD. However, we recognize that many people have different paths that can include career gaps and changes. We thus encourage anyone that feels that early career applies to them to join the organization. We also encourage those involved in the technical, operations, or engineering aspects of HEP experiments who may or may not obtain a PhD as part of their career path to join.}}
\vspace{0.2cm}

In practice, many people who fit that definition, but had already achieved permanent employment, felt that they no longer identified with the early career moniker and chose not to participate in SEC. We also had numerous discussions about the inclusion of non-scientists who work closely with scientists. In principle these people were also intended to be included in SEC, but in practice we did not see many of them in Snowmass at all, let alone in SEC. It is difficult to tell if this was due to a difference in cultural norms between fields (scientists are accustomed to a large amount of volunteer labor) or a lack of appropriate recruiting and support by the institutions which employ them. It was also unclear why early career scientists should attempt to represent engineers and technicians rather than a dedicated group doing so independently - not as a matter of exclusion, but as a matter of accurately representing each groups' interests. There was more inclusion of non-scientist voices in the Community Engagement Frontier, which addressed specific aspects of education and industry relations that required expertise from people in those fields. Non-scientists and early stage students more often struggled with some of the tools used for collaborative projects, especially Overleaf.  In order for non-scientists to be fully included in the future, there may need to be some administrative support for \LaTeX{} and document submission, if arXiv is used again. 

We defined SEC leadership to consist of the current and previous year DPF executive committee early career members and 2-3 leaders per key initiative. For the Snowmass Coordination group, each Frontier had 2-3 liaisons. The Snowmass Advisory Committee agreed that Frontier conveners should treat SEC liaisons as they would topical group conveners and include them in topical group leadership meetings as they would topical conveners. SEC leadership groups began reaching out to Frontier conveners as their leadership structure was defined. Some of the Frontiers were very welcoming of SEC participation and others were extremely opposed to it. The inclusion of SEC liaisons was an ongoing point of advocacy for the the key initiative leadership and DPF ECEC members to the Steering Committee, and the Steering Committee tended to administer their decisions through a consensus style of leadership that included the Frontier Conveners, leaving some of the SEC liaisons in the difficult position of being elected to a role they could not fulfil the duties of. In the future, it would be helpful if these kinds of decisions were made prior to the start of organizing for LOI writing and in a more binding way so that all of the participants know what to expect throughout the process. 

\section{Key Initiatives and Original Goals}
Wide community input on the goals of SEC was solicited, and five key initiatives emerged. The key initiatives included:
\begin{enumerate}
    \item In-reach
    \item Diversity, Equity, and Inclusion (DEI)
    \item Survey
    \item Long-Term Organization
    \item Snowmass Coordination
\end{enumerate}

All of these key-initiatives were planned to engage with the Snowmass process with the goal being that 1-4 would remain as a standing HEP early career representative body post-Snowmass. These are later referred to as the ``core initiatives," and yes, the distinction between `key' and `core' was always a bit strange, but that was the official distinction. The initial goals for each group were ambitious and are outlined below. These changed and evolved over time based on the resources available.

\subsection{In-reach}
In-reach was tasked with professional development and building cohesion within the early career community. Their original goals were: arranging meetings with funding agencies for early career members; organizing an informational session on the Snowmass process and how early career members could engage in the process; a workshop on letters of interest; organizing an educational series in conjunction with the Frontiers to help early career members better understand fields outside their own and better engage with Snowmass process; organizing networking opportunities at Snowmass/APS events; tracking the impact of COVID-19 on careers; and pushing to make meetings and opportunities accessible for colleagues around the world. All of these goals were Snowmass-wide, and focused on recruiting, engaging and facilitating early career participation in Snowmass generally, and throughout HEP insomuch as Snowmass is an opportunity to form connections in a concentrated environment. Inreach leadership was a good opportunity to meet a lot of people and have your name in a lot of public places, but not in a way that would emphasize your area of expertise.

\subsection{Diversity, Equity, and Inclusion}
The DEI group was charged with working on initiatives to make the HEP community representative, welcoming, inclusive, and equitable to all. Their original goals were: providing guidance to funding agencies through the Snowmass process on codes of conduct, including minority serving institutions in collaborations, and diversity, equity, and inclusion initiatives; promoting equity and inclusion in the Snowmass process and SEC leadership; developing and monitoring an anonymous feedback form; providing input on the formation of the DPF Ethics Committee; coordinating efforts with the D\&I topical group (which meant this core initiative was also a liaison group for the Diversity \& Inclusion Topical Group in the Community Engagement Frontier); encouraging work-life balance; pushing for inclusivity for engineers and technicians who are typically left out of early career organizations; and pushing to make meetings more accessible for those needing special accommodations. Leadership in this group tended to be a way to express both compassion and leadership. Leaders in this group were often the first point of contact for people who needed either the the moderators or the ethics committee just because our presence was better publicized. Since people without training should attempt to avoid receiving private details from people in difficult situations if there is another alternative, we found it necessary to link those relevant resources whenever we advertised our events or programs, and that seemed to help direct people to where they wanted to be. With some foresight, we would have done that from the very beginning and promotion of resources would have been one of our initial goals.

\subsection{Survey}
The survey group was tasked with carrying out a research survey of Snowmass members' opinions and human experiences within the field of high-energy and astrophysics. Their original goals included collecting demographic information about SEC, gathering what areas/projects were most exciting to SEC; measuring how included early career members felt in the process by Frontier; conducting a climate survey; measuring the impact of COVID-19 on early career members; and gathering community input through surveys. Distinct from previous years, this survey was for the whole Snowmass Community, not just for SEC, despite the name, which referred to the group which orchestrated it. 

\subsection{Long-Term Organization}
Long-term organization was tasked with defining the long-term structure of an early career organization that could persist after the close of the Snowmass process. Their goals included defining a structure and the continuity of SEC beyond the Snowmass process; determining how to have continuity in leadership as early career members aged; developing a webpage and channels for communication post-Snowmass; coordinating with collaboration early career organizations; and considering who in the community needed representation.

\subsection{Snowmass Coordination}
Snowmass Coordination was tasked with coordinating with the Snowmass Frontiers to help get early career members involved in the Snowmass process. Their initial goals were to attend the key meetings in their Frontiers and interact with Frontier leadership; help push key SEC initiatives within the Frontiers; present ideas on behalf of early career members in meetings if they are not comfortable presenting them themselves; helping early career members engage with their Frontiers; and sitting on major planning meetings. Each of the Coordination groups set up their more directed goals and organization independently, although the ability for leadership to meet and identify broad challenges or inhomogeneous treatment was valuable to SEC's collective advocacy influence.

\section{SEC Leadership Definition Process}
There were over 250 nominees for SEC leadership through a process that was broadly advertised. While self-nominations were allowed, many people did not know who had nominated them when they received their acceptance emails. Nominees were invited to sign up for key initiatives they wanted in either an “active participation” or “leadership” role. We then broke each key initiative into their own subgroups, made slack channels for each group, and had kick-off meetings with each of them to hand leadership over to the groups. The key initiatives functioned as their own subgroups of SEC and defined their own leadership structure. Leadership plans and current leaders were documented in a google spreadsheet. The Snowmass Coordination group decided that a Frontier-by-Frontier approach would be easier (as more than 140 nominees were involved). The group provided 2 volunteer points of contact for each Frontier to act as temporary liaisons while the leadership was defined. The activity of self-organizing in addition to setting up complicated leadership structures, all before most of us had a firm understanding of our collective priorities, was a difficult and drawn out process that resulted in the loss of many of the nominees. It was also not helpful that this process occurred much later than the selection of conveners, meaning that SEC was expected to be a fully functioning machine, able to field questions about early career priorities on specific topics or help out other groups, before many of us had a firm footing in what was going on.  There is now extensive documentation on leadership structures for different circumstances, and a strong suggestion to adopt some version of that at least to get started, which should help to streamline the process in the future.  

Most groups converged to $\sim$3 month terms with 2-3 leaders at a time. The leadership was typically chosen to be rotating terms with staggered terms to provide continuity (with only one leader cycling off at a time), but this also meant that the leadership changed roughly every month. This typically allowed all interested nominees to serve in leadership positions without elections and allowed those interested in leadership to pick terms where they would have the time to commit to these roles. This tended to work well in the beginning, when there was overall high engagement and people who were signed up to lead next were active participants throughout regardless of their official role. Later on (and keeping in mind that Snowmass 2021 was extended by a year due to COVID-19) it became more difficult to keep track of future leaders who had signed up a long time before and not kept up with the group's activities. One solution is longer terms and another is only opening the sign up for the next round, and not further and requiring current participation to sign up for the next term. Of course, were the process not extended by a year, it is unclear if such measures would be necessary. 

There were two regular monthly meetings across SEC organized by the current and previous DPF Executive Committee Early Career members. One was for Snowmass Coordination where the SEC liaisons would meet and discuss challenges, best methods for engaging their members and conveners, and identify collective goals in advocating for SEC inclusion in aspects of the work of Snowmass organizing or large events. The second was among the core initiatives to coordinate efforts and prioritize near-term goals.  These meetings often resulted in points for the core initiatives to address through event organizing or specific projects to support common issues encountered by early career participants, or messages to be carried by the DPF representatives into the steering committee meetings. It is worth noting that in practice, there tended to be a lot of overlap between the leaders of coordination groups and the leaders of core initiatives, which helped to facilitate communication between those different groups. 

\section{Maturation of Goals and Implemented Initiatives}

In this section, we describe the most active period of SEC, during which there was high generally high engagement, and a plethora of projects were undertaken to benefit the early career community and the broader Snowmass community. We highlight activities that had high impact and were successful as well as work done to set the foundation for future early career organizations for Snowmass. A lot of the activities focused on building an ethical organization from the ground up, providing career development and networking opportunities in a variety of contexts, and the construction of an advertising machine helped to promote SEC within Snowmass and Snowmass activities within and beyond those already engaged in it. 

\subsection{In-reach}
Inreach was originally understood as an early career oriented service to the community. So, we took the input from the meetings with all of the SEC nominees and tried to condense it down to a set of priorities that would best serve the community we had and to recruit additional early career people into Snowmass and make sure they knew where to go to get plugged in. 

\paragraph*{Leadership Structure} When we started, there were a lot of participants and a strong sense of inclusion behind the leadership structure. The system we designed had 2 staggered leader roles, each lasting 4 months. There was a blind signup form with months of availability and one volunteer put the schedule together trying to make sure that everyone that wanted to could serve as leader during the year. Then, the full draft and process were revealed. This was an attempt to reduce bias toward strong personalities in a 'first come-first served' style signup. Although, it is possible to force the system by only providing availability for the first term. 

In practice, the leaders were the point of contact for other areas of Snowmass and those responsible for generating zoom links and updating the meeting calendar and notes documents. Inreach was otherwise a group of project leaders. Each person able took on one project to lead: monthly colloquia, special event series, networking coffee meetups, etc. After a few switches, this system started to break down because some people who had not been leaders early on would wait until their term came up to attend the meetings or participate at all, leading to a lot of catching up work. Additionally, since early career people change life circumstance/jobs so frequently, many people who signed up for the end of the year were unavailable when their turn came up. All of this was understandable on a personal level, but I mention it because it demonstrates a flaw in the system we built. The decay looked something like people serving multiple terms and then eventually no leadership remaining, at which point Inreach merged with DEI and LTO to form the ``SEC Core Initiatives" which carried on the Colloquium series, and one final recruiting push after the end of the 6-month break due to the pandemic. 

\paragraph*{Group \& Project Organization} The weekly meetings were used to keep other members of the group up to date on upcoming efforts, balance the schedule to make sure we were not overproducing content for our audience, and to gather extra helpers for any time sensitive tasks that came up as projects progressed. This model allowed us to work independently, reducing the need for constant collaboration, but also provided a support structure. Since a project without someone to lead it was not one on our roster, this setup also helped to keep projects the group took on at a manageable level for participants. One drawback, was that since everything going on had the main personnel it needed most of the time, it could be difficult for newcomers to figure out how to join in - it took initiative to join the group and participate actively even though the intent was not to be exclusive. There was very much a sense that we wanted our events to have an impact and that as such they should not be so frequent that the average Snowmass participant was unable to keep up with them. It was understood that people would move on and projects without a lead would die and that was fine. It is important to keep the transient nature of Snowmass in mind when considering the time individual early career people devote to it because it can be intense and small tasks seem to take on out sized importance when you are in the middle of them with passionate people. 

\paragraph*{Early Events} We organized a one-hour Introduction to Snowmass for Early Careers event that defined the Snowmass process for people who had not been part of one before. There was a lot of interest among the younger people and a lot of questions that were otherwise receiving speculative answers in slack could be addressed authoritatively all at once. We had a speaker who was senior in the field and well versed in Snowmass as well as a panel of informed early career representatives who were able to answer questions. 

Following that successful event, we organized a series of Frontier Introductions. This was a 3 part series with 3 Frontiers per part (one Frontier was invited, but did not participate). Each part was 90 min long to allow for 20 min presentation and 10 min discussion from each Frontier, and they were spaced at 1 week intervals. We invited the Frontier conveners to describe the topics that fit within their Frontier, the exciting science they wanted to highlight, and how especially early career people could get plugged into the Frontier activities and working groups for letters of interest and later white papers. 

\paragraph*{LOI \& Paper Workshops} We had extensive discussions about hosting a letter of interest mini-workshop which would have been an hour or so lecture describing the structure and purpose of the letters and what to include to make them effective since the writing style is different than for publications.  This would also have been a good chance to describe Snowmass procedures surrounding the LOIs. However, we did not have the personnel to organize it at the time it would have been needed, and there was some hesitancy to explain a concept that some people thought was too simple to warrant a presentation. 

Some similarly nebulous discussion of a white paper mini-workshop was also floated, but there were even fewer Inreach participants, some white papers were already underway, the break was approaching, meaning few people would be working on papers imminently anyway, and when we came back from the break, several months later than originally discussed, the focus for Snowmass participants needed to be on organizing their science white papers and between lack of man power and concern for being a distraction, the Core Initiatives also did not take up that challenge. There was also some confidence that most people remaining after the break had integrated into their working groups already and any newcomers could be pointed to existing groups that would be able to better explain the specifics. It is unclear if the absence of the pandemic would have presented a more obvious point in time when such an event would have been helpful to a large number of people (noting that undergraduate and early graduate students were generally not participants in Snowmass from our early demographic surveys).

\paragraph*{Tips for Successful Meetings} We found that Fridays between 12pm and 2pm Eastern Time were the best compromise for timing because it is not too late in Europe and not too early in California. Other days of the week people tend to be thinking about other things and on Fridays more people are open to sitting in on a talk or two. Broad advertising via multiple methods is important to reach as many people as possible. The event was advertised on Slack, Twitter (through the SEC account), and via email listservs to the general and snowmass-young groups.  We recorded the zoom meeting and placed the video on YouTube.  

Future event planners should be mindful of trying to have large events live-captioned to make them accessible to people who have a hard time listening to the call, which may arise from deafness, English as a second language, noisy or quiet environment for the listener, etc. The steering committee did not provide funding for this service and did not address multiple requests for it. So, we always provided captions in our Youtube uploads, but of course this prevents active participation from people who would otherwise have attended in real time. 

\paragraph*{Big Questions in Particle Physics Colloquium Series} By far the most consistent and successful program that Inreach ran was the monthly colloquium series. \href{https://www.youtube.com/channel/UCFzOxX28tsZ32_j43oe5XFw/videos}{Youtube}\footnote{\href{https://www.youtube.com/channel/UCFzOxX28tsZ32_j43oe5XFw/videos}{https://www.youtube.com/channel/UCFzOxX28tsZ32\_j43oe5XFw/videos}} videos of these events are available. We tried to focus each event either on one Frontier from 2 different perspectives: theory \& experiment, earth \& cosmic, etc, OR on one broad topic with overlap in at least 2 Frontiers each of which could send a representative. Speaker selection was key to a good program because they often chose to coordinate with each other and the unusual style of talk meant they could not use something prepared for a different event. Speakers tended to be mid- to late-career with broad expertise in the topic of interest.  

We asked them to provide the context of the topic from their perspective for an audience of scientists outside of their field, since the series was for all career stages across all Snowmass Frontiers. Then, we asked them to identify the big open questions that they though would or should be addressed by the community over the next 50 years in this area. Each speaker was allotted 25 min + 5 min for questions on their individual presentation, then 30 min were reserved for a broader discussion with the audience about the topic and open questions. In practice, this 30 min period was also a useful buffer for long-winded talks, extra questions, and anything else that might come up during a live event. 

When the event was hosted by Inreach, the host was generally the main organizer from the Inreach group, which changed a few times. When the event was hosted by SEC Core Initiatives, we tended to rotate based on availability and expertise since the event organization process was much more collaborative. There were extensive discussion about some form of this series continuing after Snowmass ends, but featuring early career scientists and narrower, but still future looking topics. In the end, the Core Initiatives leadership decided to host the final event in June, to finalize the series prior to the Community Summer Study in July. With almost all of the members in transitional phases of their careers, it was a bad time to try to recruit and re-brand to keep it going in addition to a general sense that Snowmass activities had gone on for too long. 

\subsection{Diversity, Equity, and Inclusion}

The Snowmass Early Career - Diversity, Equity, and Inclusion (SEC-DEI) group focused on promoting DEI within SEC and to the broader community through Snowmass related activities.   

\paragraph*{Leadership Structure}
When we started, there were a lot of participants and a strong sense of inclusion behind the leadership structure. The system we designed had 3 staggered leader roles, each lasting 6 months. We used a Google Sheet as the ongoing sign-up form and the system was 'first come-first served.' The leaders were the point of contact for other areas of Snowmass and those responsible for generating zoom links and updating the meeting calendar and notes documents. Additionally, the SEC-DEI leaders served as the discussion moderators during the meetings. For much of the active period of the group, SEC-DEI leaders met once during the week to discuss projects, issues, and plan the meeting. Then, there would be another meeting hosted for the full group of participants. Occasionally, the full group were invited to the leadership meeting if something of broader (but not full group) importance needed an update in between full meetings. 

It was not intended that the SEC-DEI rotating leaders were also necessarily the liaisons to the D\&I Topical Group, but that was often the case since the leaders were more engaged during their term than the average member. It was also not intended to be a requirement that the leaders took on a majority of the project lead positions, but that was also often the case. There was a lot of interest in DEI among early career people who did not have time for extensive involvement, and there is nothing in principle wrong with peripheral involvement - we each do what we can and manage our own schedules to the best of our ability, having periods of higher or lower participation that tend to balance as a community. However, the system of the full group voting on what projects the group should undertake, rather than volunteers taking up projects they could manage, tended to result in the leadership feeling responsible for an unreasonable and unsustainable amount of work. This is not an uncommon problem DEI as a broader subfield - people lead by compassion tend to feel guilt for not helping whether they have the bandwidth to help or not. We hope that future groups are more mindful of this particular pitfall and the potential for burnout of overextended leadership, especially in compassionate work. 

\paragraph*{Topical Group Liaison Role}
During some of the first meetings of the group, there was a vote of those in attendance for this group (one of the four core initiatives) to also serve as the liaising group to the Diversity \& Inclusion Topical Group (CEF03) in the Community Engagement Frontier. Notably, the rest of the Community Engagement Frontier was liaised through the SEC-Community Engagement (SEC-CEF) Coordination group. This cross-roads in relevance was by happenstance and the vote likely reflected either a general misunderstanding about the intended structure of SEC or a desire among the participants in attendance to be more involved with fewer meetings. This dual role contributed to some additional responsibilities being placed on fewer people, and confusion about `which hat' a leader was wearing when speaking to some issue was not uncommon. In the future, we would recommend maintaining logical divisions between groups with distinct purposes, while encouraging coordinated participation. For example, the two groups could have met on alternating weeks during the same time block. 

\paragraph*{Early Efforts}
One of the early goals assigned to SEC-DEI was to provide a Code of Conduct for SEC. After some discussion about pulling resources from various universities, national labs, and societies, we realized there was a parallel effort going on to create the ``Core Principles \& Community Guidelines (CP\&CG)\footnote{\href{https://drive.google.com/file/d/1dE1-Qm4iZfzWVo8L97Cpz1UnOjOkQ2mm/view}{https://drive.google.com/file/d/1dE1-Qm4iZfzWVo8L97Cpz1UnOjOkQ2mm/view}}" for all of Snowmass. The leaders of SEC-DEI met with the Ethics Task Force (which later became the DPF Ethics Committee), offered comments based on our recent discussions and determined that there was no need for SEC-DEI to continue to invest time into an effort that was already in capable hands elsewhere in Snowmass. The CP\&CG are now a component of DPF and apply to all DPF-sponsored meetings, so there should be no future need for early career people to independently revisit the idea of creating a Code of Conduct, although as a `living document' any DPF member should be able to comment on the current version of the CP\&CG if it is found lacking in the future. 

Another idea that came up early on for SEC-DEI was to set up an anonymous dropbox. We took the names of volunteers (the leaders), set up an anonymous GoogleForm explaining the purpose and who would have access to the responses. The purpose was intended to be low level fielding of community concerns that we could then advocate for without anyone knowing who had requested the change. In constructing the reader's team, we wanted 3 people with different demographics along as many axes as possible to reduce bias and blind spots. In practice, we did not get many responses. Of those that did arrive, they tended to be small things like complaints about the community posting information in the wrong slack channel, which were both out of our control and not difficult to post a little bit of guidance about to help point people toward the channels they needed. That we advertised the dropbox probably contributed to people contacting the leadership directly about more serious problems they encountered either in Snowmass or at their institution, which we were not prepared to deal with. Nevertheless, we tried to point people toward the resources they needed, either in the moderator channel or to the Ethics Committee. In the future, it should probably be the role of the Ethics Committee to set up such a general dropbox given the unpredictable nature of the comments, and their more official role. However, a future SEC-DEI could request that such a dropbox be set up to receive feedback on the Snowmass process while it is going on, and that there be a topical group whose explicit responsibility is to receive feedback on the Snowmass Process because several different communities complained about how disenfranchised they felt at various points, and that kind of information is probably relevant to the success of Snowmass in representing all of the communities relevant for the offices at DOE and NSF-Physics that it claims to represent. 

The last major undertaking in the early stages of SEC-DEI was to provide accountability and support for diversity within the SEC organization. We interpreted that as specifically trying to understand the inclusion efforts in leadership structures that were put in place, and whether those systems had resulted in leader diversity along various axes. It was our intention to observe, support, and eventually report in aggregate, but not to interfere since we did not interpret our role as one of power over our colleagues. To this end, we set up a demographic survey for SEC leaders, specifying how the information would be used, providing opt-out by not requiring answers to any of the questions except for name, email, and group where they served as a leader. The readers (again lacking alternate volunteers, the SEC-DEI leaders) were specified on the form and sworn to secrecy apart from the discussions. After the questionnaires were returned, it was clear that some groups had refused to participate by filling the survey, but not answering any of the questions. We transposed the results to anonymize them for the intended long term record of diversity in SEC leadership. We set up meetings with individual leadership groups to discuss how they tried to incorporate principles of inclusion in both their leadership structures and how the groups were run. We intended to offer tips like, keeping some positions open for new people, or making sure that meetings have notes and make space for less assertive personalities. However, when we actually met with each of the groups that wanted to meet with us, they tended to have their own set of concerns, like low participation, the relevance of Snowmass in their careers, or antagonistic conveners. These types of conversations helped us to better advocate for SEC in more anonymized ways, and know who to check in with from time to time about how especially the smaller groups were doing. It also helped us to realize that collaboration with Inreach on recruiting efforts fit without our core goals. If we had this to do over, the idea of checking on leadership demographics is not necessarily bad, but in the context of untrained people doing demographic work with names and small groups, it was probably too invasive. It was also a lot of work, and not necessary to the discussions that were actually helpful, which we could have had just by checking in with each of the leader groups one on one to see how things were going and where we could pull resources to help. 

Even though, we would not recommend an unsupervised demographic survey of our peers, the process of thinking and discussing through how to best perform a demographic survey ethically resulted in a \href{https://docs.google.com/document/d/e/2PACX-1vTU17wmMpieO2v3CezXWclFYdLxol_waxDL4KyGIGymRtImXn98jd0M2NY5AsjsWD_0jml5WURGFjU7/pub}{Letter of Interest}. At the core of those discussions was how to construct demographic categories inclusively, while maintaining the ability to report results without revealing an individual identity. There was a sense of resentment against the U.S. Census categories, specifically for race and ethnicity because the groupings do not represent most minorities well. We also had extensive discussions about making sure that participants can give informed consent, and strictly treating the survey description as a contract between the participants and the specified survey team, which cannot be changed by one party without the consent of the other.

\paragraph*{Advocating for Accessibility}
SEC-DEI, and persons in leadership roles in other areas of Snowmass who also participated in SEC-DEI, were involved in a lot of the early recruiting and informational meetings, as well as large, Snowmass-wide meetings throughout the process. As such, it was not uncommon for SEC-DEI leadership to receive requests for, and complaints about the lack of, accessibility services for our events and Snowmass events more generally. 

The Snowmass 2021 process took place over Zoom, with the exception of the Community Summer Study. Scientists and students who are deaf, hard of hearing, spoke English as a second language, had to attend meetings from a loud environment, and those who needed to attend the meeting from an environment where they could not make noise, would all have benefit from live captioning. At the time of this Snowmass, artificial intelligence captioning services tended to be in beta. They tended to be easily accessible from University Zoom accounts, but had an average failure rate of about 1 in 6 words during normal conversation, for speakers with a good connection and no accent. The failure rate tended to be worse for speakers with accents, people who enunciate less than average, or for any speaker discussing technical terms or using acronyms. The software tended to place it's best guess rather than alert the reader that something might be off, which often resulted in unintelligible transcripts. Google AI captioning for YouTube after the fact tended to be better, but live AI captioning was a constant sticking point in arguments about accessibility because members of the Deaf community said they may as well not attend events using it, and Snowmass leadership could not be persuaded away from the notion that it was a reasonable compromise/interim solution. 

The SEC Core Initiatives petitioned the Steering Committee through the DPF EC Representatives at their meetings several times and failed to make headway. The group then set up a meeting with the Ethics Committee to ask them to petition the Steering Committee for a solution to accessibility at Snowmass meetings, especially those for Early Career participants. There was an issue of funding for any captioning service since DPF's budget is not increased in the year they host Snowmass.  There were issues just in the game of telephone that information had to pass through so many people to get from a complainant to someone capable of action that key information was always lost, there was little accountability for a response.  

Eventually a small amount of grant funding ($\sim\$60$k) was procured to provide live captioning for some Frontiers to have them at their last big meetings before the CSS. The funding was allocated to the Frontiers who indicated a desire for support for those meetings, but they were not informed that the support was intended for accessibility services. The Steering Committee decided that none of that funding would be allocated to any SEC events, and never provided any notice or reasoning for that decision. Some accessibility services were provided at the CSS.  

While we hope that AI captioning improves significantly over the next decade and this issue in particular will not pose such a significant barrier to future participants, there is something to be said for the creation of real-time feedback pathways that are efficient, responsive, and accountable.  Additionally, to host such a massive community driven process requires money. We strongly recommend that DPF budget for accessibility, as well as administrative support for each Frontier, as part of their pre-Snowmass planning in the future, and procure grants or donations in advance to support these elements of the program. 

\paragraph*{Journal Club}
During a lull in SEC-DEI projects that involved a lot of planning, the participants were looking for a way to be more engaged and there had been a lot of discussion about making use of Physics Education and Sociology literature, rather than just `feelings' or personal experiences of those present to provide justification for DEI actions within Snowmass and at our home institutions. As a result, there was a vote to create a journal club and hold it on alternating weeks with the business meetings. We set up a spreadsheet for signups and made use of a dropbox for article suggestions. However, each week when a volunteer was needed to present an article, there were none. For about 4 sessions, the leadership tried to introduce some overviews and demonstrate how to find and present journal club articles, but were not willing to continue the program as the sole presenters indefinitely. After some backlash, the Journal Club was placed on indefinite hiatus pending volunteer presenters and simply never returned. It takes a lot of active participation to keep a journal club going, which tends to require a driving force, like enrollment in a class, or someone of stature running it. Interest in a topic is not sufficient to establish a journal club. Additionally, in the context of Snowmass, it is not clear that this activity would have served the central mission of SEC-DEI - be mindful of scope-creep. People interested in DEI topics beyond the scope of Snowmass could be introduced to organizations like APS-IDEA, AAAS, or other advocacy networks as a more productive and longer lasting outlet for those interests. Remember that Snowmass is a pop-up organization, and there are well established advocacy organizations with much to share with early career particle physicists. 

\paragraph*{Developing Inclusive Organizational Recommendations}
SEC-DEI devoted several months to the development and refinement of a set of recommendations for SEC leadership. Most of these recommendations came from observing how groups had self-organized, what worked for them, and what lead to pitfalls later on. This was heavily informed by our discussions with other SEC leaders as well as the time period in which they were developed. 

The recommendations \href{https://docs.google.com/document/d/e/2PACX-1vRJSi2VP_BXd3GnKVpme1hE9X6yIYFBEHNP4T-skM7mhxYJctMsBMCnzMzvOJQF7CFZp1NZKVYDWNvh/pub}{(Link to Recommendations)} developed a system that attempted to remove as much bias as possible, while still allowing for broad participation in the selection of leaders. There was a tension between the idea that people introduce bias and that people should feel that they are part of the process of choosing their leaders. We hope that elections would not result in various forms of tribalism, especially choosing only individuals from large collaborations or from prestigious institutions. We also hoped that overall there would be a mix of genders, ethnoracial groups, sexualities, socioeconomic backgrounds, career stages, and where relevant specialities. However, we recognized that there was a need to trust future nomination and voting processes to be informed and take the principles of inclusion to heart in their personal decision making processes. 

With the possibility for less-savory characters coming into leadership positions, especially positions of public trust like SEC-DEI, we did want the ability to vet candidates, but for privacy and practicality, we knew that peer vetting was going to be untenable. We consulted with the DPF Ethics Committee who agreed they had the ability to assist with future elections (although, reminding them in 10 or so years will probably be necessary due to personnel turnover). The idea is that they get a list of nominees, the community is informed of those nominees so that they have the opportunity to provide confidential feedback to the Ethics Committee. If they don't hear anything, they just return the list. If they do hear complaints, they can follow up to the best of their ability and use their judgement about whether the nominee they received a complaint about should be eligible for the position. To maintain confidentiality, the ethics committee only returns the result to the election committee, either approved or not approved. We envision that in most cases, nothing will happen during the vetting period, but we felt it was important to create a reporting avenue anyway. 

There is a standard set of instructions for running the elections and a slightly simplified set for small groups. The thinking was that in simplifying the instructions for smaller groups we reduce the administrative burden on a group of people that would naturally be more accountable to each other due to the nature of smaller group dynamics. Basically trying not to over-complicate situations where the leaders are the only participants because there just are not many participants. However, we do recommend revising the version of the rules a group follows if they grow or shrink past the threshold. 

\paragraph*{Early Career Experience Survey \& Panel Discussions}
SEC-DEI organized a large survey focused on the experience of early career particle physicists and astrophysicists. Some of the questions focused on interactions between the individual and other people grouped by career stage, with the rising idea that bias from peers and subordinates can also take a toll, even though conversations tend to focus more on biases of senior colleagues. We looked at how ones relationship with their dissertation or postdoc advisors can shape their career outcomes, how perception of being overworked tracked with hours spent working, and how that changed with career stage. We asked about whether people knew about and took advantage of various career development milestones, like review panels, publication, committees and workshops. We asked about experiences with harassment and how it was addressed in different ways depending on how it was reported. Notably, conferences hosted by professional societies tend to produce the best outcomes for victims in terms of addressing the problem and providing support. There were some topics we asked about because we thought they would be important and there was no statistically significant correlation. There were correlations we had not expected. While in retrospect, we would not recommend repeating such an intrusive survey of our peers without professional guidance, the results were pivotal in informing the construction of the panel discussions for early career people on DEI topics. In particular, it helped to remove us as organizers from the microcosm of a self-selected DEI group and actually address the concerns of the community we purported to serve. 

The Panel Discussions that we organized in response to this survey were promoted alongside another series organized by the D\&I topical group. One of the recordings is available on YouTube\footnote{\href{https://www.youtube.com/channel/UCFzOxX28tsZ32_j43oe5XFw/videos}{https://www.youtube.com/channel/UCFzOxX28tsZ32\_j43oe5XFw/videos}}, alongside other recordings of SEC-Inreach events. The series was very successful, and we would recommend facilitating similar discussions for Early Career people as a matter of making the field more broadly accessible and reducing barriers created by assumed/inside knowledge.

\subsection{Survey}
\paragraph{Successes}
The Snowmass Early Career Survey Initiative has been a far reaching and successful venture within the Snowmass Process. Entirely lead by graduate student and postdoctoral level scientists, this small but highly active group has endeavored to serve as a model for inclusivity and respectful free inquiry among our colleagues, investigating our mutual community's collective dynamics, shortcomings, and aspirations. The Survey, initially conceived within the 2001 Snowmass Process~\citep{Fleming:2001zk} and further developed within the 2013 Process \citep{Anderson:2013fca} has been similarly updated and expanded for the present 2021 Process. Topics within the Survey included demographics, career outlook, physics outlook, the effects of the COVID-19 pandemic, along with workplace and social climates throughout high-energy and astrophysics. Unlike some previous Survey Initiatives~\citep{Fleming:2001zk,Anderson:2013fca}, the 2021 Survey did not limit itself to querying only the Early Career membership of the Process, but instead the whole of the community; this decision was made by the Survey team after months of ongoing discussions and codevelopment of survey topics and questions. The Survey received more than 1500 interactions, and a far-reaching white paper will be submitted to the Community Engagement Frontier showing and contextualizing the results. The full report is available~\citep{Agarwal:2022gno}, and will be updated gradually as more data is analyzed.


\paragraph{Challenges}
One of the biggest challenges faced by the Survey team were arriving at consensus among the group in how to approach difficult, emotionally and perhaps politically-charged topics, including harassment and racism (or at least racial disparities) in physics. Great care was needed to investigate these questions as they pertain to the overall climate within the field, and due diligence was paid to their development. Key to these and other question developments was a small but highly devoted set of reviewers who perused, discussed, and took the Survey in its entirety before full distribution; this group included past Snowmass Survey team members (notably from the 2013 report~\citep{Anderson:2013fca}) and the SEC DEI Initiative.

Another challenge was quite simple: time and effort. In retrospect, it is hard to imagine how the Survey team would have been able to put out the report given the previous Snowmass timeline; the extension provided us with ample time for extra discussion, question development, analysis, and interpretive writing. However, it should be noted that this extended time period saw many team members change jobs, most moving from graduate to postdoctoral work, and thus limiting their time for further development, analysis and writing. Key to any future successful Survey will be a core group of perhaps 5-10 early career members who can commit to such work over a potentially longer-haul than initially expected.

\subsection{Long-term Organization}
While a long-term organization was not established this Snowmass, the long-term organization group identified and investigated several leadership structures that could be pursued in the future. Two primary ideas emerged. One is developing and expanding a network of existing early career organizations. Another would be forming an advisory group to the APS DPF executive committee.

\paragraph{An Early Career Network:} One idea that seemed promising for a future early career organization in HEP was developing a network of existing early career groups (e.g. the Fermilab Student and Postdoc Organization, Young NO$\nu$A, Young DUNE, Young Mu2e, Young CMS, etc.). In our collective experience, such groups act as excellent tools to inform and empower early career scientists and provide them with a platform to promote their interests and needs. These organizations make it easier for EC members to contribute and have their work recognized, greatly enabling networking and connection opportunities across and outside the field. Maintaining such self-led representational groups within management, publication, and conference committees removes career barriers, promotes equity and inclusion, offers professional development opportunities, and gives EC scientists a voice among their more established peers. A network between these early career groups currently does not exist.

The envisioned network could take the form of two points of contact from each EC organization. Information and opportunities gathered across the broader HEP community would be a valuable resource, especially to those without current early career representation and those seeking to establish early career groups of their own. This network of contacts could be maintained by the early career member(s) on the DPF executive committee. Such a network could: (1) Establish and communicate best practices among existing early career groups, improving the efficacy of sharing ideas across organizations; (2) Inspire and support early career members who do not have an organizations to develop them; (3) Extend the benefits and opportunities of an early career organization to communities where such organizations are difficult to develop, including colleagues working in theory or in small-scale collaborations; (4) Facilitate ambitious coordinated initiatives and opportunities that may have been previously out of reach for individual early career groups; and (5) Enable policies of mutual respect, inclusivity, equity, justice, human dignity, and freedom of inquiry to be more easily universalized and supported across the field.

\paragraph{Significant Challenges}
While several ideas were developed and gained traction, one of the primary issues with developing a long-term organization was the need for sustained leadership. Recruiting and retaining leadership can be challenging, especially when early career positions can be short-lived. It was not clear how an early career organization could continue without a strong plan for recruitment and retention, so this would need more development should an early career organization be pursued in the future.

It was also unclear how to balance the focus of the early career organization since graduate students, postdocs, and early faculty/scientists all have very different needs. 

\paragraph{APS DPF Advisory Committee:}
Another idea explored in detail was developing an APS DPF Early Career advisory committee. Such a committee could consist of the early career DPF Executive Committee representatives, the leadership of the key initiatives, and some number of board members. Membership in the early career organization would be open to all in the field. 

Leadership Committee terms would be two years with an option to continue two years more, and terms could be staggered to foster continuity. Since the committee would be via APS, committee members would need to be members of APS. There were concerns that this could exclude those who cannot fund the membership fees, so there was a push to advertise resources for funding fees and potential discussion of fee waivers for committee members (though this would require negotiation with APS). The positions on the board would include a Chair, Deputy Chair, DPFEC Liaison, DPFEC Advisor, Administrator, Treasurer, Webmaster, and Subcommittee/Initiative positions. The current DPF Executive Committee early career member would hold the DPF Liaison role, while the previous year's early career member would hold the role of DPFEC Advisor. The Chair, Deputy Chair, and Subcommittee Chairs would be elected by the Leadership Committee, while Webmaster, Treasurer, Administrator would be filled on a volunteer-basis or named by the Chair. 

Nominations for Leadership Committee would be open to the community, and the current Leadership committee would appoint the new committee members according to membership guidelines. The appointment of each member would be decided upon at a meeting of the Leadership Committee and by majority $2/3$ vote from the Leadership Committee. The Leadership Committee would appoint incoming members considering: nominee statements, number of nominations, diversity of fields, and general diversity. Egregious concerns about a candidate could be brought to the Chair and one other committee member, who could voice this concern and/or veto the nominee.

New subcommittees and/or initiatives could be added by petition from at least three Leadership Committee members and majority vote of the Leadership Committee. At least one of the petitioners would chair/co-chair the initiative/subcommittee until the following calendar year, designated at the time of proposing the sub-committee. Subcommittees/initiatives could be removed by procedural review with a yearly vote applying to each subcommittee before chair elections (a 2/3 majority would be required for removal). Bylaws could be amended by discussion of amendments by petition of at least two Leadership Committee members at a meeting of the Leadership Committee and previous notice in writing one week before a meeting of the Leadership Committee. Amendments would require discussion time, followed by a 2/3 vote of the Leadership Committee. Unsuccessful amendments could be re-proposed after a period of at least one week.

This proposed structure would need further community input and collaboration with the APS DPF committee to be an advisory committee.

\subsection{SEC Coordination}
SEC Coordination describes the network of Early Career Liaisons to specific Frontiers. Per the instruction of the Steering Committee, SEC-Representatives (limited to 2-3 per meeting) should be included in any meeting or communication as if they were representing a topical group for that Frontier. In practice, there was a wide variety of inclusion and exclusion of early career leadership and participants, which the collective advocacy power of SEC was not sufficient to address since we lacked sufficient representation in the Steering Committee who were generally unwilling to enforce their decisions on conveners who volunteered their time. 

Since the SEC Frontier liaisons were implemented later in the process, some Frontier conveners were reluctant to include early career liaisons in their meetings. Additionally, some Frontiers felt that they should have had the power to appoint their SEC liaisons, while SEC felt that we should be able to chose our own representation. Some liaisons misunderstood their charge and thought that they were required to attend every meeting versus being able to attend, so some SEC coordination groups ballooned in size, exceeding 2-3 liaisons, and having 2-3 liaisons per topical group. There was a general lack of consensus about the appropriate role of Liaisons. 

At a high level, liaisons were intended to sit in on Frontier meetings, pass information to their constituent SEC-[Frontier] group, gather input, and pass consensus input from the subfield specific Early Career Community back to the relevant Conveners in order to include early career perspectives in the planning process. 

SEC-Energy was an especially successful example of a coordination group that rotated leadership, maintained high activity, and coordinated well with their conveners. SEC-Theory was an example of an organization with good intentions that struggled with engagement due to a hostile environment among more senior participants. SEC-Cosmic struggled to be included in meaningful decisions early on and spent much of the process in an administrative and advertising role. While those activities were pivotal to the larger success of the Snowmass process, they should have been undertaken by a paid administrative assistant. SEC-Community Engagement, tended to be so well included in every activity that they were eventually just absorbed by the main groups, including several early career people who stepped into convener roles positions opened later in the extended process. 

In the future, we recommend an earlier nomination period and a more uniform approach to including early career representatives. While SEC provided the personnel and organizational structure, it is clear in retrospect that the successes of the organization were largely driven by the cultures of the Frontier organizations we came alongside.

\bibliographystyle{style}    
\bibliography{main}   

\end{document}